\documentclass[prl,twocolumn,showpacs,preprintnumbers,amsmath,amssymb,floatfix,longbibliography]{revtex4-1}

\usepackage[pdftex]{graphicx}
\usepackage{epstopdf} 
\usepackage{graphicx}
\usepackage{dcolumn}
\usepackage{color} 
\usepackage{bm}

\usepackage[colorlinks]{hyperref}
\hypersetup{
	pdfstartview={FitH},
	linkcolor=blue,
	citecolor=blue,
	filecolor=blue,
	urlcolor=blue
}

\begin{document}

\title{Enhancement of Spin Currents in a Lattice near the Curie Point}
	
\author{Marco Finazzi}
\email{marco.finazzi@fisi.polimi.it}
\affiliation{LNESS-Physics Department, Politecnico di Milano, Piazza Leonardo da Vinci 32, 20133 Milano, Italy}
\author{Federico Bottegoni} \author{Carlo Zucchetti} \author{Giovanni Isella} \author{Franco Ciccacci} 
\affiliation{LNESS-Physics Department, Politecnico di Milano, Piazza Leonardo da Vinci 32, 20133 Milano, Italy}
\date{\today}

\begin{abstract}
	We show that pure spin currents carried by conduction electrons injected into a paramagnetic lattice of mutually interacting localized magnetic moments can be enhanced  close to the Curie temperature by the exchange interaction between the lattice sites and the non vanishing spin density associated with the spin current. The latter partially aligns the magnetic moments of the lattice, generating a flow of paramagnons that contribute to the total spin current. 
\end{abstract}

\maketitle

	
The active control and manipulation of the spin degree of freedom in solid-state systems is the goal of a branch of electronics known as spintronics, which has its foundations on the investigation of spin transport, dynamics and relaxation in electronic materials \cite{Wolf2001,Zutic2004}. Spintronics has already demonstrated a huge potential in consumer electronics and has stirred a technological revolution in the field of mass storage media. Spin-based devices that have already made their way to the market include giant and tunnel magneto-resistance reading heads for hard drives and non-volatile magnetic random-access memories (MRAM) \cite{Akerman2005}, where information is stored in the parallel or antiparallel alignment between two ferromagnetic (FM) layers.
	
As much as charge currents are the key ingredients of electronics, spin-polarized currents are essential in spintronics. A spin-polarized current carries an angular momentum that can be transferred to a FM layer, inducing oscillations of its magnetization or even flipping it completely \cite{Slonczewski1996}. This phenomenon, known as \textit{spin-transfer torque} (STT), can be exploited for encoding information in MRAMs in a way that holds the promise of being faster, cheaper and less power-hungry than other existing solutions \cite{Kent2015}. However, the fast switching of devices relying on this encoding mechanism still requires current densities that are too high for most commercial applications. Therefore, a strategic issue in spintronics consists in finding solutions to increase the total spin that can be transferred by an electron current.
	
In the following, we evidence how a pure spin current due to carriers flowing in a lattice of disordered (paramagnetic) localized magnetic moments can be much larger than the sum of the spins individually associated with each free particle, if the material is kept at a temperature close to the Curie point. In these conditions, the lattice magnetic susceptibility is considerably enhanced as a consequence of the large spin fluctuations and, thanks to the exchange interaction, the spin accumulation associated with the spin current of mobile carriers results in a sizable magnetization of the lattice. This allows elementary lattice excitations known as paramagnons to diffuse in the solid and contribute to the total spin current.

Such a phenomenon might suggest strategies to emphasize spin-dependent phenomena in solids, leading to more efficient spintronic assets, such as faster STT devices or effective architectures for the interconversion between spin and charge currents. 
	
\section{\label{Discussion} Discussion}

\subsection{\label{sub1} Spin currents in a paramagnetic ensemble of local moments}

Let us consider a lattice of atoms carrying disordered local magnetic moments, such as the ones associated with the $d$ electrons in a paramagnetic transition metal \cite{Izuyama1963}, in interaction with $s$-derived spin-polarized conduction electrons \cite{Takahashi2011}.
Our goal is to estimate the expectation value $\langle\hat{J}^S_{\alpha\beta}\rangle$, with ${\alpha, \beta=x,y,z}$, of the spin-current density operator defined as \cite{Bottegoni2012}:  
\begin{equation}\label{eq:1}
	\hat{J}^S_{\alpha\beta}(\mathbf{r},t)=\frac{e}{m_e}\mathrm{Re}
	\left[ \hat{\psi}^\dagger(\mathbf{r},t)\,\sigma_\alpha\,\hat{p}_\beta\,\hat{\psi}(\mathbf{r},t) \right], 
\end{equation}
where $e/m_e$ is the electron charge/mass ratio, $\sigma_\alpha$ one of the Pauli matrices, and $\hat{p}_\beta$ the momentum operator projected along the $\beta$ axis,
\begin{equation}\label{eq:2}
	\hat{p}_\beta=\hbar\sum_{\mathbf{k},\varsigma}k_\beta \, \hat{a}^\dagger_{\mathbf{k},\varsigma}(t)\ \hat{a}_{\mathbf{k},\varsigma}(t).
\end{equation}
In Eq.~(\ref{eq:1}),  $\hat{\psi}^\dagger(\mathbf{r})$ and $\hat{\psi}(\mathbf{r})$ are the creation and destruction field operators, respectively, for the ensemble of local moments:
\begin{subequations}\label{eq:3}
	\begin{equation}\label{eq:3a}
		\hat{\psi}^\dagger(\mathbf{r},t) = \frac{1}{\sqrt{V}} \sum_{\mathbf{k},\varsigma}\hat{a}^\dagger_{\mathbf{k},\varsigma}(t) \, e^{i\mathbf{k}\cdot\mathbf{r}},
	\end{equation}
	\begin{equation}\label{eq:3b}
		\hat{\psi}(\mathbf{r},t) = \frac{1}{\sqrt{V}}\sum_{\mathbf{k},\varsigma}\hat{a}_{\mathbf{k},\varsigma}(t) \, e^{-i \mathbf{k}\cdot\mathbf{r}},
	\end{equation}
\end{subequations}
$\hat{a}^\dagger_{\mathbf{k},\varsigma}$ and $\hat{a}_{\mathbf{k},\varsigma}$ being the creation and destruction operators in momentum representation, respectively, with $V$ indicating the sample volume and $\varsigma$ the quantum number corresponding to the spin projection along the chosen quantization axis.

The coupling between the localized and conduction electrons is assumed to be a local $s$-$d$ exchange interaction \cite{Takahashi2011}:
\begin{align}\label{eq:4}
	\hat{\mathcal{H}}_{sd}(t)&=\frac{\mathcal{J}_{sd}}{N}\int_V \mathbf{s}(\mathbf{r},t)\cdot\hat{\mathbf{S}}(\mathbf{r},t)\, d\mathbf{r}\nonumber\\
	&=\frac{\mathcal{J}_{sd}}{N}\sum_\mathbf{k} \mathbf{s}^*(\mathbf{k},t)\cdot\hat{\mathbf{S}}(\mathbf{k},t),
\end{align}
$N$ being the number of lattice cells per unit volume and $\mathcal{J}_{sd}$ the coupling constant, with ${\mathcal{J}_{sd} < 0}$ since the exchange interaction always favors an anti-parallel alignment between the magnetic moment localized on the lattice sites and the spin of the conduction electrons \cite{Yosida1996}.

In Eq.~(\ref{eq:4}), $\mathbf{s}(\mathbf{r},t)$ and $\mathbf{s}(\mathbf{k},t)$ represent the spin density (in units of $\hbar$) associated at time $t$ with conduction electrons and its spatial Fourier transform, respectively. In the following we will consider $\mathbf{s}(\mathbf{r},t)$ as a small external perturbation, due to spin-polarized carriers injected in the paramagnetic lattice. Interactions between the local moments mediated by the conduction electrons, such as the Rudermann-Kittel-Kasuya-Yosida (RKKY) interaction \cite{Yosida1996}, will be treated implicitly by considering the appropriate unperturbed Hamiltonian (see the next section). In Eq.~(\ref{eq:4}), $\hat{\mathbf{S}}(\mathbf{r},t)$ and $\hat{\mathbf{S}}(\mathbf{k},t)$ are the spin density operators associated with the local orbitals and its spatial Fourier transform, whose Cartesian components are respectively defined as:
\begin{subequations}\label{eq:5}
	\begin{equation}\label{eq:5a}
		\hat{S}_\gamma(\mathbf{r},t)= \hat{\psi}^\dagger(\mathbf{r},t)\, \sigma_\gamma\,\hat{\psi}(\mathbf{r},t),
	\end{equation}
	\begin{equation}\label{eq:5b}
		\hat{S}_\gamma({\mathbf{k},t)}= \frac{1}{\sqrt{V}} \sum_{\mathbf{q},\varsigma,\varsigma'}\hat{a}^\dagger_{\mathbf{q},\varsigma}(t)\,\sigma_\gamma \,\hat{a}_{\mathbf{q}+\mathbf{k},\varsigma'}(t),
	\end{equation}
\end{subequations}
with $\gamma=x,y,z$.

The expectation value of the double Fourier transform of the spin current can be obtained from the Kubo formula \cite{Mahan2000}:
\begin{align}\label{eq:6}
	&J^S_{\alpha\beta}(\mathbf{k},\omega) =\langle{\hat{J}}^S_{\alpha\beta}(\mathbf{k},\omega)\rangle =-i\frac{\mathcal{J}_{sd}}{\hbar N} \sum_{\gamma} s^*_\gamma(\mathbf{k},\omega) \nonumber\\
	&\times  \lim_{\varepsilon\to 0}\int_{0}^\infty \left\langle \left[\hat{J}^S_{\alpha\beta}(-\mathbf{k},t), \hat{S}_\gamma(\mathbf{k},0)\right]\right\rangle_0 e^{-i\omega t-\varepsilon t}\, dt,
\end{align}
where $s_\gamma(\mathbf{k},\omega)$ is the double Fourier transform of $s_\gamma(\mathbf{r},t)$. The bracket in the integral indicates the equilibrium average of the unperturbed system, defined as $\langle\hat{A}\rangle_0=\mathrm{Tr}[\hat{\rho}\hat{A}]/\mathrm{Tr}\left[\hat{\rho}\right]$, with $\hat{\rho}$ being the density matrix describing the thermodynamic properties of the ensemble of local moments in the absence of the perturbation $\mathbf{s}\left(\mathbf{r},t\right)$.
In the previous expression, the spin current density operator $\hat{J}^S_{\alpha\beta}(\mathbf{k},t)$ is obtained from Eq.~(\ref{eq:1}) and takes the following form:
\begin{align}\label{eq:7}
	&\hat{J}^S_{\alpha\beta}(\mathbf{k},t) = \frac{e\hbar}{2m_e \sqrt{V}}\sum_{\mathbf{q},\varsigma,\varsigma'} \left(k_\beta+2q_\beta\right) \hat{a}^\dagger_{\mathbf{q},\varsigma}(t)\,\sigma_\alpha\,\hat{a}_{\mathbf{q}+\mathbf{k},\varsigma'}(t)\nonumber\\
	&= \mu_B \left[k_\beta\hat{S}_\alpha(\mathbf{k},t) + \frac{2}{\sqrt{V}} \sum_{\mathbf{q},\varsigma,\varsigma'} q_\beta\, \hat{a}^\dagger_{\mathbf{q},\varsigma}(t)\,\sigma_\alpha\,\hat{a}_{\mathbf{q}+\mathbf{k},\varsigma'}(t)\right],
\end{align}
$\mu_{B}$ being the Bohr magneton. At this point one should notice that, in the long wavelength limit (${k\rightarrow 0}$), which applies when the driving term $\mathbf{s}\left(\mathbf{r}\right)$ vary slowly over many lattice sites, the first term in the right-hand side of Eq.~(\ref{eq:7}) is \textit{even} with respect to time-reversal, described by the following substitutions: $\hat{a}^\dagger_{\mathbf{q},\varsigma}\rightarrow\hat{a}^\dagger_{-\mathbf{q},-\varsigma}$, $\hat{a}_{\mathbf{q},\varsigma}\rightarrow\hat{a}_{-\mathbf{q},-\varsigma}$, and $\sigma_\alpha\rightarrow -\sigma_\alpha$. Conversely, the second term in the right-hand side of Eq.~(\ref{eq:7}) is \textit{odd}. Since a spin current should be \textit{even} upon time-reversal, the contribution of the second term to $\langle{\hat{J}}^S_{\alpha\beta}(\mathbf{k},\omega)\rangle$ should vanish in the long wavelength regime and only the first term, the one proportional to $\hat{S}_\alpha$, needs to be considered. Its substitution into Eq.~(\ref{eq:6}) yields:
\begin{align}\label{eq:8}
	J^S_{\alpha\beta}(\mathbf{k},\omega) & = \mu_B k_\beta\langle \hat{S}_\alpha(\mathbf{k},\omega)\rangle  \nonumber\\
	&=\mu_B k_\beta\sum_{\gamma}  \frac{\mathcal{J}_{sd}}{N} \chi_{\alpha\gamma}(\mathbf{k},\omega) \, s^*_\gamma(\mathbf{k},\omega) \nonumber\\
		&=-i
	\mu_B k_\beta\sum_{\gamma}  \frac{\mathcal{J}_{sd}}{N} \chi_{\alpha\gamma}(\mathbf{k},i \omega_n) \, s^*_\gamma(\mathbf{k},\omega) \nonumber\\
		&=-i\mu_B k_\beta\sum_{\gamma} \eta_{\alpha\gamma}(\mathbf{k},i \omega_n) \, s^*_\gamma(\mathbf{k},\omega),
\end{align}
where $\chi_{\alpha\gamma}(\omega)$ is the magnetic susceptibility of the lattice of local moments, defined as in Ref.~\cite{Izuyama1963}:
\begin{align}\label{eq:9}
	&\chi_{\alpha\gamma}(\mathbf{k},\omega) = \nonumber\\
	&= -\frac{i}{\hbar} \lim_{\varepsilon\to 0}\int_{0}^\infty \left\langle \left[\hat{S}_\alpha(-\mathbf{k},t), \hat{S}_\gamma(\mathbf{k},0)\right]\right\rangle_0 e^{-i\omega t-\varepsilon t}\, dt,
\end{align}
and $\eta_{\alpha\gamma}$ is the tensor that describes the linear relation between the vectors $\langle\hat{\mathbf{S}}\rangle$ and $\mathbf{s}$. Following Ref.~\cite{Mahan2000}, in Eq.~(\ref{eq:8}) we have indicated with $\chi_{\alpha\gamma}(\mathbf{k},i \omega_n)$ the analytic continuation of $\chi_{\alpha\gamma}(\mathbf{k},\omega)$ obtained for $i\omega_n\rightarrow\omega+i\delta$. 
The bracket in the integral of the previous expression corresponds to the spin density correlation function \cite{Izuyama1963}, describing the spatial and time fluctuations of the spin density in the unperturbed material.

In Eq.~(\ref{eq:8}), the term $-ik_\beta s_\gamma(\mathbf{k},\omega)$ corresponds to the spatial Fourier transform of $\partial_\beta s_\gamma(\mathbf{r},\omega)$, which is associated with a spin diffusion current density $j^s_{\gamma\beta}$ carried by conduction electrons. In a paramagnetic medium the latter assumes the following form \cite{Fabian2010}:
\begin{equation}\label{eq:10}
	j^s_{\gamma\beta}(\mathbf{r},\omega) = \frac{\partial_\beta s_\gamma(\mathbf{r},\omega)}{e\rho(\omega)D(\epsilon_F)},
\end{equation}
where $\rho(\omega)$ is the resistivity and $D(\epsilon_F)$ the density of states at the Fermi energy $\epsilon_F$, with $D(\epsilon_F) =\frac{3}{2}\frac{N_e}{\epsilon_F}$ within a free-electron model where $N_e$ is the density of conduction electrons. Note that the term $j^s_{\gamma\beta}$ reported in Eq.~(\ref{eq:10}) would be the only contribution to a \textit{pure} spin current density, consisting in a flow of spins with no charge transport, a condition that requires a null electric field.  By substituting the expression of $j^s_{\gamma\beta}$ into Eq.~(\ref{eq:8}) one obtains:
\begin{equation}\label{eq:11}
	J^S_{\alpha\beta}(\mathbf{k},\omega) = \frac{3}{2}\frac{N_e}{\epsilon_F}e\rho(\omega)\mu_B \sum_{\gamma} i\,\eta_{\alpha\gamma}(\mathbf{k},\omega) \, j^s_{\gamma\beta}(\mathbf{k},\omega).
\end{equation}
Assuming typical values for transition metals, $N_e\approx {10^{29}\;\mathrm{m}^{-3}}$, $\epsilon_F=\frac{\hbar^2}{2 m_e}\left(3\pi^2 N_e\right)^\frac{2}{3}\approx 7\;\mathrm{eV}$, $\rho\approx {10^{-7}\;\Omega}$m, one obtains that the adimensional pre-factor multiplying the sum appearing in Eq.~(\ref{eq:11}) is about ${2\times 10^{-2}}$.

The previous result can be interpreted as follows: in compounds characterized by long-range magnetic (ferromagnetic or anti-ferromagnetic) order, small oscillations of the moments around their equilibrium direction will propagate as waves, called magnons. In a paramagnetic material, the magnetic order is partially re-established by the spin accumulation $\mathbf{s}(\mathbf{r})$, which polarizes the lattice of local moments thanks to the exchange interaction, allowing for spin waves to propagate. These waves, known as \textit{paramagnons}, undergo diffusive transport and can contribute to the total spin current. The concept was first proposed by Berk and Schrieffer \cite{Berk1966}  and Doniach and Engelsberg \cite{Doniach1966} to explain additional repulsion between electrons in some metals, which reduces the critical temperature for superconductivity.

An equally legitimate interpretation consists in viewing the total spin current as due to quasiparticles known as \textit{spin polarons}, formed by a conduction electron dressed in a cloud of lattice moments aligned with the electron spin by the exchange interaction \cite{Storchak2009a}. Because of the paramagnon group velocity being much smaller than the Fermi velocity, the total lattice spin dragged by each conduction electron would just be a fraction (expressed by the pre-factor in Eq.~(\ref{eq:10})) of the amount of spin that would surround a \textit{static} spin such as, for instance, the one associated with a magnetic impurity. Indeed, the formation of spin polarons typically leads to strong electron localization \cite{Storchak2011}.

\subsection{\label{sub2} Enhancement of a pure spin current}

From the conclusions of the previous section, one can expect a significant enhancement of a spin-diffusion current promoted by the paramagnetic lattice of local moments in materials endowed by large magnetic susceptibilities. In this respect, the best candidates would be the so-called \textit{nearly-ferromagnetic} materials, which just fail to satisfy the Stoner criterion for the onset of long-range magnetic order. Typical examples are represented by the near-noble metals Pd and Pt, which develop ``giant moments'' around diluted magnetic impurities such as Fe or Co atoms. In these systems a single magnetic impurity may polarize up to 200 neighboring host atoms \cite{Low1966} and induce a magnetic moment as large as 12~$\mu_B$ per Fe impurity in Pd or 6~$\mu_B$ per Fe impurity in Pt \cite{Crangle1965}. This is possible because the density of conduction electrons at the Fermi level is sufficiently low to allow the existence of sharp impurity levels associated with high unscreened magnetic moments \cite{Clogston1962}, which polarize the $d$ levels of the surrounding atoms of the Pd or Pt matrix through the RKKY interaction. 
Another example is MnSi, which is a weak ferromagnet with $T_\mathrm{C}=29.5$~K and is, from the magnetic point of view, at the borderline between an itinerant electron metal and a system of local moments. In this compound, strong electron localization into a bound state suggested to be a spin polaron has been found in both paramagnetic and ferromagnetic states, with a net spin ${S = 24\pm2}$ per electron \cite{Storchak2011}.

The susceptibility of such materials will be calculated in the long-wavelenght, quasi-static regime characterized by ${k\rightarrow0}$, ${\omega_n\rightarrow0}$ \cite{Mahan2000}. We consider an isotropic and homogeneous system (which implies ${\eta_{\alpha\gamma}=\eta\,\delta_{\alpha\gamma}}$) described by the Heisenberg Hamiltonian with the addition of the perturbing term $\hat{\mathcal{H}}'_{sd}$ given by Eq.~(\ref{eq:4}): 
\begin{equation}\label{eq:12}
	\hat{\mathcal{H}} = -\sum_{i,j}\mathcal{J}_{dd}(\mathbf{r}_i-\mathbf{r}_j)\;\mathbf{S}_i\cdot\mathbf{S}_j+\mathcal{J}_{sd}\sum_j\mathbf{S}_j\cdot\mathbf{s}\left(\mathbf{r}_j\right),
\end{equation}
with $\mathbf{S}_j$ indicating the spin (in units of $\hbar$) localized on the $j^{\mathrm{th}}$ lattice site and $\mathcal{J}_{dd}\left(\mathbf{r}_i-\mathbf{r}_j\right)$ the exchange coupling between sites $i$ and $j$.

In the random phase approximation (RPA) \cite{Izuyama1963}, the average value $\left\langle\mathbf{S}_i\right\rangle$ of the spin at the $i^{\mathrm{th}}$ lattice site at temperature $T$ is obtained by simultaneously solving the following set of equations \cite{Izuyama1963, Blundell2007}:
\begin{subequations}\label{eq:13}
	\begin{equation}\label{subeq:13a}
		\left\langle S_i\right\rangle/S = \mathcal{B}_J\!\left[\frac{g\mu_{B}J\left|\mathbf{B}_{sd}(\mathbf{r}_i)+\mathbf{B}_{dd}(\mathbf{r}_i)\right|}{k_B T}\right];
	\end{equation}
	\begin{equation}
		\mathbf{B}_{sd}(\mathbf{r}_i) = \frac{\mathcal{J}_{sd}}{g\mu_{B}}\mathbf{s}\left(\mathbf{r}_i\right);
	\end{equation}
	\begin{equation}\label{subeq:13b}
		\mathbf{B}_{dd}(\mathbf{r}_i) = -\frac{2}{g\mu_{B}}\sum_j\mathcal{J}_{dd}(\mathbf{r}_i-\mathbf{r}_j)\left\langle\mathbf{S}_j\right\rangle.
	\end{equation}
\end{subequations}
In Eq.~(\ref{subeq:13a}), $\mathcal{B}_J$ is the Brillouin function for a total angular momentum quantum number $J$. Because of the quenching of the angular momentum due to the crystal field, in the following we will assume ${J \approx S}$, being $S$ the total spin quantum number. $\mathbf{B}_{sd}$ and $\mathbf{B}_{dd}$ are the effective magnetic fields associated with the $s$-$d$ and $d$-$d$ exchange coupling, respectively, and $g\approx2$ is the electron gyromagnetic ratio. The factor 2 in Eq.~(\ref{subeq:13b}) appears because of  the double counting in the first sum of Eq.~(\ref{eq:12}).

The set of Eqs.~(\ref{eq:13}) is associated with a relevant energy $E_\mathrm{C}$ defined as
\begin{equation}\label{eq:14}
	E_\mathrm{C} = \frac{g\mu_{B}}{3}\lambda S\left(S+1\right),
\end{equation}
with
\begin{equation}\label{eq:15}
	\lambda =-\frac{2}{g\mu_{B}}\sum_j\mathcal{J}_{dd}(\mathbf{r}_i-\mathbf{r}_j) \approx -\frac{2N}{g\mu_{B}}\int\mathcal{J}_{dd}(\mathbf{r})\,d\mathbf{r}.
\end{equation}
$E_\mathrm{C}$ has a simple interpretation in the Weiss model \cite{Blundell2007}, which assumes a uniformly magnetized state with ${\left\langle\mathbf{S}_i\right\rangle = \left\langle\mathbf{S}\right\rangle}$ exposed to a uniform externally applied magnetic field, with $\lambda$ representing  the proportionality constant between the uniform spin $\left\langle\mathbf{S}\right\rangle$ of the system and the Weiss effective field ${\mathbf{B}_{dd} = \lambda \left\langle \mathbf{S}\right\rangle}$ describing the average $d$-$d$ exchange interactions.
In this case, for ${\lambda>0}$, Eqs.~(\ref{eq:13}) admit a uniform ferromagnetic ground state solution when $T$ is below the Curie temperature defined as ${T_\mathrm{C}=E_\mathrm{C}/k_B}$. However, one must remind that the RPA slightly overestimates the Curie temperature of a real FM system \cite{Ashcroft1976}. Therefore, in the following, $E_\mathrm{C}$ should be considered as a parameter describing the magnitude of the (average) exchange interactions between the lattice sites, rather than a quantity strictly related to the true ordering temperature of the system. However, we would also like to add that, despite the RPA is known not to be able to perfectly describe correlations close to phase transitions, it is also widely accepted as a means to capture the essential physics behind complex phenomena.

We consider solutions of Eqs.~(\ref{eq:13}) when the system is in the paramagnetic state at a temperature ${T > E_\mathrm{C}/k_B}$, approximating the solution by linearizing the Brillouin function: ${\mathcal{B}_J(x) \approx \mathcal{B}_S(x) \approx x(S+1)/(3S)}$. By applying the continuum approximation, Eqs.~(\ref{eq:13}) then yield
\begin{equation}\label{eq:16}
	\mathbf{S}(\mathbf{r}) = \frac{1}{\lambda \theta}\left[\mathbf{B}_{sd}(\mathbf{r}) -\frac{2N}{g\mu_{B}} \int \mathbf{S}(\mathbf{u})\mathcal{J}_{dd}(\mathbf{r}-\mathbf{u})\,d\mathbf{u}\right],
\end{equation}
where ${\theta = k_B T/E_\mathrm{C} > 1}$ is the normalized temperature. By Fourier-transforming Eq.~(\ref{eq:16}) one obtains the following solution:
\begin{align}\label{eq:17}
	\mathbf{S}(\mathbf{k}) &= \frac{g\mu_{B}}{g\mu_{B}\lambda \theta+2N\sqrt{V}\mathcal{J}_{dd}(\mathbf{k})}\mathbf{B}_{sd}(\mathbf{k})\nonumber\\
	&= \frac{\mathcal{J}_{sd}}{g\mu_{B}\lambda \theta+2N\sqrt{V}\mathcal{J}_{dd}(\mathbf{k})}\mathbf{s}(\mathbf{k}) = \eta(\mathbf{k})\mathbf{s}(\mathbf{k}).
\end{align}
In the long wavelength limit (${k\rightarrow 0}$), $\eta$ takes the following form:
\begin{equation}\label{eq:18}
	\eta(k\rightarrow 0) = \frac{2}{3}
	 S\left(S+1\right) \frac{\mathcal{J}_{sd}}{k_B T-E_\mathrm{C}}.
\end{equation}

In the case of conduction-electron-mediated RKKY coupling between the lattice sites, one can derive in a straightforward manner the relation between $\mathcal{J}_{sd}$ and $E_\mathrm{C}$, since $\mathcal{J}_{dd}$ is given by the following expression \cite{Yosida1996}:
\begin{equation}\label{eq:19}
	\mathcal{J}_{dd}(\mathbf{r}_i-\mathbf{r}_j) = -9\pi\frac{\mathcal{J}^2_{sd}}{\epsilon_F}\left(\frac{N_e}{N}\right)^2\mathcal{F}(2k_F\left| \mathbf{r}_i-\mathbf{r}_j\right|),
\end{equation}
where $k_F$ is the Fermi wavevector, $N_e/N$ is the number of conduction electron provided by each atom, and $\mathcal{F}\left(x\right)={(\sin x-x\cos x)x^{-4}}$. 
Combining Eqs.~(\ref{eq:14}), (\ref{eq:15}) and (\ref{eq:19}), one eventually finds
\begin{equation}\label{eq:20}
	E_\mathrm{C} = S\left(S+1\right)\frac{\mathcal{J}^2_{sd}}{\hbar \, \mu_B\left(3\pi^2 N\right)^\frac{2}{3}}\left(\frac{N_e}{N}\right)^2.
\end{equation}
Focusing our attention on Pd and Pt, where ${S\approx \frac{1}{2}}$, and $N_e \approx N \approx {10^{29}~\mathrm{m}^{-3}}$, we obtain $\mathcal{J}_{sd}\approx -2.8\sqrt{E_\mathrm{C}}$ when both $\mathcal{J}_{sd}$ and $E_\mathrm{C}$ are expressed in electron-volts. Pure Pd and Pt are paramagnetic at any temperature, but they can be turned into ferromagnets by doping with magnetic impurities such as Fe, Co or Ni \cite{Clogston1962}, with Curie temperatures that can be as high as ${T_C=104\;\mathrm{K}}$ for $6.3\%$ Fe in Pt and ${T_C=236\;\mathrm{K}}$ for $9.8\%$ Fe in Pd. We should also remind that in 4$d$ and 5$d$ metals the high spin-orbit interaction might be comparable with the crystal-field splitting, leading to a partial unquenching of the orbital moment \cite{Blundell2007}. For late transition elements such as Pd and Pt, the orbital and spin moments are parallel, meaning that there might be a small net contribution due to the orbital moment to the total angular momentum carried by the enhanced spin current.

\begin{figure}[t]
	\includegraphics[width=1\columnwidth]{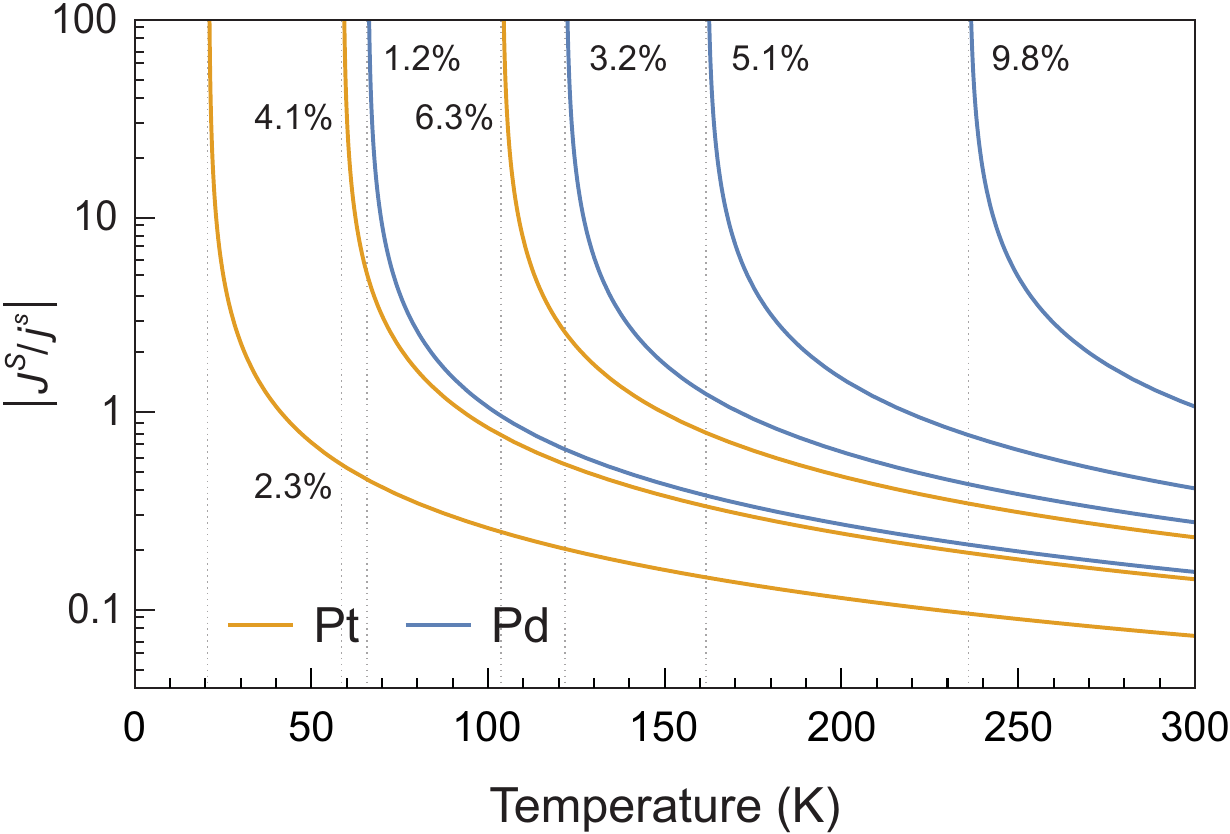}
	\caption{\label{fig:1} Temperature dependence of the absolute value of the $|J^S/j^s|$ ratio estimated for diluted Fe-Pd and Fe-Pt alloys in the paramagnetic phase. The Curie temperature $T_C$ (indicated by the vertical dotted lines) is determined by the Fe concentration, reported for each curve. The $T_C$ values have been obtained from Ref.~\cite{Crangle1960} and from Ref.~\cite{Crangle1965} for Pd- and Pt-based alloys, respectively.}
\end{figure}

Figure~\ref{fig:1} reports the temperature dependence of the absolute value of the $|J^S/j^s|$ ratio estimated for diluted Fe-Pd and Fe-Pt alloys for different values of the respective Curie temperatures, suggesting that enhancement factors of at least one order of magnitude might be within reach, albeit in a narrow temperature range close to $T_C$. This conclusion is corroborated by the observation of strong enhancements of the Spin Hall effect near the Curie point of NiPd \cite{Wei2012} and FePt \cite{Ou2018} alloys, which is attributed to large spin fluctuations, a condition that, according to Eq.~(\ref{eq:9}), leads to a large susceptibility of the paramagnetic lattice. 

The aforementioned spin current enhancement mechanism also bears strong similarities with the diffusion of thermal antiferromagnetic (AFM) magnons \cite{Khymyn2016, Rezende2016}, a phenomenon that has been proposed to explain why spin currents propagating in a AFM insulator can significantly enhance the input spin current \cite{Wang2014, Lin2016, Qiu2016}. Noteworthy, the spin current transfer through the AFM layer is found to depend on its temperature, substantially increasing when $T$ approaches the N\'{e}el temperature \cite{Khymyn2016} and showing a trend very similar to the one displayed in Fig.~\ref{fig:1} by the $|J^S/j^s|$ ratio.

This spin enhancement can only occur when carriers flow in a material kept at a temperature close to the Curie point, while spintronic devices such as MRAMs require injecting the spin current into a FM layer, where the strong exchange field strongly suppresses the fluctuations of the local moments and thus the carriers are no longer dressed with the spin excitations of the lattice. Therefore, the point is now to understand whether the spin-current enhancement provided by a paramagnet could be exploited in a device.
	
Let us consider the interface between a FM and a NM layer. Let $z$ be the axis perpendicular to the interface, with the FM layer at ${z>0}$ and the NM one at ${z<0}$. Let us assume a pure spin current is traveling parallel to $z$ in the NM layer. It is worth noticing that, when NM = Pt and FM = Fe, the interface spin resistances of the two metals, $r^\mathrm{Pt} = \rho^\mathrm{Pt} L_s^\mathrm{Pt}$  and $r^\mathrm{Fe} = \rho^\mathrm{Fe} L_s^\mathrm{Fe}$ \cite{Fert2001} (with $\rho$ and $L_s$ corresponding to the resistivity and spin diffusion length, respectively) are quite similar to each other \cite{Rojas2014, Bottegoni2015}, thus preventing possible issues related to the conductivity mismatch for the transfer of spin-polarized electrons across the Pt-Fe interface \cite{Fert2001}. Moreover, interface NM atoms are magnetically coupled to the FM material, allowing paramagnons diffusing in the NM layer to transfer angular momentum to the FM one, where magnons are excited, and to contribute to the STT \cite{Yan2011}. 

The total angular momentum deposited per unit time into the slab is equal to the total spin current multiplied by $\hbar/(2e)$ \cite{Slonczewski1996, Vanhaverbeke2007}. Despite  spin scattering mechanisms could transfer part of the angular momentum to the lattice, the generation of spin-torque by magnons is a well-documented phenomenon \cite{Qiao2018, Demidov2010, Madami2011, Demidov2012, Hamadeh2014}. Therefore the injection of paramagnons into the FM material is expected to significantly contribute to the STT when the system is kept close to the NM Curie temperature.

\section{\label{Conclusions} Conclusions}

We have shown that a pure spin current flowing into a lattice of disordered magnetic moments might be enhanced by the exchange interaction that aligns the magnetic moments of the lattice with the spin of the carriers. The spin-current enhancement is inversely proportional to the difference between the system temperature and the critical temperature describing the strength of the exchange interaction among lattice sites, and becomes significant when the two differ by a few kelvins.



%

\end{document}